\documentclass[10pt,aps,prb,twocolumn,noshowpacs]{revtex4-1}
\usepackage{amsmath,graphicx,amsbsy,amssymb,amsmath,epsfig,latexsym,wasysym,pifont,mathrsfs,natbib}
\usepackage{float} \newfloat{widefig}{thp}{lop} \usepackage{comment}
\usepackage{array, makecell} \usepackage{verbatim} \usepackage{datetime}
\usepackage{multirow,array} \usepackage{color} \usepackage{setspace}
\usepackage{subfig}
\usepackage{hyperref} 
\hypersetup{colorlinks,
  linkcolor=blue,%
  citecolor=blue,%
  urlcolor=blue}

\begin{document}

\title{Systematic investigation of capacitances in functionalized MXene supercapacitors $M_{n+1}C_nO_2$, $M=Ti,V,Nb,Mo$}
\author{Mandira Das}
\email[]{mandira@iitg.ac.in}
\affiliation{Department of Physics, Indian Institute of Technology
  Guwahati, Guwahati-781039, Assam, India.}    
\author{Subhradip Ghosh}
\email{subhra@iitg.ac.in} \affiliation{Department of Physics,
  Indian Institute of Technology Guwahati, Guwahati-781039, Assam,
  India.} 

\begin{abstract}
MXene, the class of two-dimensional materials, has been found to be useful as potential electrode materials for electrochemical capacitors. Although experimental investigation on the electrochemical performances of a few MXenes have been carried out with exciting results, a complete understanding of their atomic scale behaviour is yet to be done. Using first-principles electronic structure methods, we perform a systematic investigation of the capacitances in pristine and functionalised MXenes $M_{n+1}C_{n}O_{2}$ where $M=Ti,V,Nb$ and $Mo$. We provide results on each of the three sources of the capacitance and analyse them in detail for a complete understanding of their behaviour. The interpretation of the experimental results, wherever available, in the light of our computations,provides useful insights.   

\end{abstract}

\maketitle

\section{Introduction}\label{introduction}
Of late, energy storage devices that may provide the dream combination of high energy density, high power density and long life cycle have been the subject of intense investigation \cite{wang2008supercapacitor}. Among them,the supercapacitors have attracted substantial attention because of its  wider scope of  applications in cases where more energy output is necessary within a short time. This is because of the high power density and long cycle life of supercapacitors in comparison with conventional batteries\cite{ren2017overview,masaki2019hierarchical,yang2020applications}. The performance of a supercapacitor crucially depends on it's capacitance value and the amount of charge it can store. The charge storage in supercapacitors can be either through a surface redox reaction or through electrostatic attraction. The former leads to pseudocapacitance while the later one leads to Electrical double layer capacitance (EDLC) which is way higher than the capacitance of an usual dielectric capacitor. The contribution to the total capacitance in a supercapacitor comes from both leading to the net electrical capacitance ($C_E$) being the resultant capacitance due to parallel combination of pseudo-capacitance and EDLC. A third important contribution to the total capacitance comes from quantum capacitance, $C_Q$,a property of the electrode material. It originates due to the inability of the metal electrodes to perfectly screen the electric field. Due to it's presence the total capacitance decreases since $C_Q$ is connected in series with the total electrical part of the capacitance \cite{lury}. Being very large, $C_Q$ is irrelevant for most materials except for nanosystems where it is found to affect the overall capacitance significantly \cite{mead,saad} since the total capacitance is given as
\begin{equation}
\frac{1}{C_T} = \frac{1}{C_Q} + \frac{1}{C_E}
\label{Eqn:1}
\end{equation}
 
 With higher surface to volume ratio, two-dimensional nanomaterials are ideal as supercapacitor electrode materials as the availability of large surface area enable them to store large amount of charge. Expectedly,Graphene emerged as a promising electrode material\cite{yu2015three} for supercapacitor. Researchers also explored carbon-nanomaterials\cite{chen2018ultrasmall}, metal-oxides\cite{zhang2014porous}, hydroxides\cite{gao2014ultrahigh}, metal dichalcogenides\cite{xu2018direct} as alternatives. In 2011, with the discovery of $Ti_3C_2$\cite{naguib2011two}, the 2D family had a new class of nanomaterials, MXenes, with chemical composition $M_{n+1}X_{n}$; $M$ a transition metal, $X$ either Carbon or Nitrogen, $n$ an integer. In last decade, more than 30 MXenes\cite{khazaei2019recent} have been synthesised and they turned out to exhibit multiple functionalities.Like other 2D nanomaterials, many MXenes have emerged as promising electrode material for supercapacitors. The first reported capacitance value was as large as  $900F/cm^{-3}$ at a scan rate of $2mV/s$ in 1M $H_2SO_4$\cite{ghidiu2014conductive} electrolyte was for $Ti_3C_2$. This value of capacitance was substantially higher than that achieved in Graphene. Subsequently, $Mo_2C$, and $V_2C$ were also explored. While $Mo_2CT_x$ showed a gravimetric capacitance of $190F/g$ in 1M $H_2SO_4$ solution\cite{halim2016synthesis} $V_2CT_x$ registered a capacitance of $487F/g$ at scan rate of $2mV/s$ in same electrolyte solution\cite{shan2018two}. 
 
In spite of demonstrating their potential as supercapacitor electrodes, very little has been done to understand the origin of supercapacitance in MXenes at the eletronic level. Theoretically, efforts have been spent in understanding the pseudo-capacitance in $Ti_{3}C_{2}$ only \cite{zhan2018,wang2019}. Therefore, in this paper,we have done a systematic investigation of various contributions to the total capacitances of six MXene compounds $Ti_2C$, $V_2C$, $Nb_2C$, $Mo_2C$, $Ti_3C_2$ \& $Nb_3C_2$ using first-principles Density functional theory (DFT) \cite{dft} based methods. The reasons behind choosing these six are the following: (i) all the MXene compounds except $Nb_3C_2$ have been synthesised and (ii) Whether the trends in their capacitative behaviour can be traced back to the behaviour of the $d$ electrons of the transition metal constituents. This is the reason for selecting compounds with transition metal component from among both $3d$ and $4d$ elements in periodic table. MXene is a 2D derivative of its 3D precursor MAX phase compound. Passivation of it's surface by functional groups like fluorine, oxygen, hydroxyl group is inevitable in the process of synthesising MXene from the corresponding MAX phase. This surface passivation can significantly affect the material properties. Although a number of robust theoretical calculations have been performed on different MXenes, many of them represent idealised situation by considering pristine MXenes only. This runs the risk of missing out important physical effects due to surface passivation. Thus,in this work,we have considered both pristine and oxygen functionalised systems. To our knowledge, no such systematic and comprehensive study addressing the behaviour of various components that make up the total capacitances in MXenes is available till date. Our calculations demonstrate that (i) all compounds considered have reasonable minimum values of the capacitances (ii) consideration of $C_Q$ is essential to have a realistic estimate of the total capacitance and (iii) there is no notable trend with regard to the $d$ orbitals of the transition metal $M$.   
 \section{Details of calculation method}\label{methods}
 We have used DFT based electronic structure method to obtain the information at the electronic level and used them to compute various contributions towards total capacitance. In the following sub-sections we describe the approaches considered for computing the capacitances.
 
 \subsection{Quantum Capacitance}\label{QC}
 Quantum capacitance of a material depends upon its electronic structure. As the capacitor charges, the negative plate fills up with electrons, which occupy higher-energy states in the band structure, while the positive plate loses electrons, leaving behind electrons with lower-energy states in the band structure. Therefore, as the capacitor charges or discharges, the voltage changes at a different rate than the electrical potential difference.In these situations, One must  consider the band-filling effect, related to the density-of-states (DOS) of the plates, for calculation of the total capacitance. The most significant changes in the DOS are  near the Fermi level. Thus, Quantum capacitance $C_Q$ is given as
\begin{equation}
C_Q = \frac{dQ}{dV}
\label{Eqn:2}
\end{equation}
where $Q$ and $V$ are the charge and voltages respectively. The variation in the total charge around the Fermi level due to application of voltage $V$ is expressed as,
\begin{equation}
\Delta Q = e \int_{-\infty}^{+\infty} D(\xi)[f(\xi)-f(\xi-eV)] d\xi
\label{Eqn:3}
\end{equation}
where $D(\xi)$ and $f(\xi)$ are the density of states of the material and Fermi-Dirac distribution respectively.
The final expression for differential quantum capacitance is given by,
\begin{equation}
{C_Q}^{diff} = \frac{dQ}{dV} = e^2\frac{dN}{dU}= e^2\int_{-\infty}^{+\infty}D(\xi)F_T(\xi-U)d\xi
\label{Eqn:4}
\end{equation}
$N$ is the total number of electrons and $U$ the chemical potential.$F_T(\xi-U)$ is the thermal broadening function,given by,
\begin{equation}
F_T(\xi-U)=\frac{1}{4k_bT}sech^2(\frac{\xi-U}{2k_bT})
\label{Eqn:5}
\end{equation} 
The important quantity that gives a clear picture of a material's storage capacity is integrated quantum capacitance($C_Q^{int}$). $C_Q^{int}$ is the material's charge storage capacity when it is charged up to a certain voltage.
$C_Q^{int}$ is given as,
\begin{equation}
\label{Eqn:6}
{C_Q}^{int} = \frac{1}{eV}\int_{0}^{V} {C_Q}^{diff}dV
\end{equation}
Total stored charge $Q$,therefore, is,
\begin{equation}
Q = C_Q^{int}V
\label{Eqn:7}
\end{equation}
The DOS obtained from self-consistent electronic structure calculations is used in Equations \ref{Eqn:4}-\ref{Eqn:6} for calculations of Quantum capacitance. In all our calculations, the temperature is kept at 300K.  
\subsection{Pseudo-Capacitance \& EDLC}\label{PSC}
The pseudo-capacitance is the electrode material's charge storage capacity when ions accumulate over the electrode's surface and charge transfer happens across the double layer. Theoretically,the pseudo-capacitance $C_{redox}$ can be obtained either from the slope of the integrated DOS of the transition metal(qualitative estimation)\cite{zhang2016} or by computation of the ratio of transferred charge ($\Delta Q$) and voltage change($\Delta V$) \cite{kang2011,zhan2018}.In either method, the calculations were done considering a neutral electrode. This may lead to missing out of important effects as in reality electrodes store and release charges during the operation of the capacitor. Recently Wang {\it et al} \cite{wang2019} used the second method to calculate pseudocapacitance of $Ti_3C_2T_2$ by considering the electrode in a charged condition. They adopted the model of hydrogen ion adsorption in the electrode to simulate the process of interaction between a negatively charged electrode and positively charged hydrogen ions of the electrolyte in a double-layer capacitor. In order to calculate $C_{redox}$, $\Delta Q$ was considered to be the electron number gained by the adsorped hydrogen ions while $\Delta V$ during the charging was computed from the change of the work function using the relation $\Delta V = \Delta WF/e$; $\Delta WF$ the change in the work function due to hydrogen ion adsorption, $e$ is the elementary charge of electron. This model uses the fact that upon adding or releasing electrons from the system, the band structure undergoes only a rigid shift. Under this so called 'Rigid band approximation' (RBA), the only effect on the electronic structure is the shift in the position of the Fermi level upon changes in number of electrons. 

Accordingly, the work function change after hydrogen ion adsoption on the electrode is defined as,
\begin{equation}
\Delta WF = WF - (WF^{neutral} - \Delta E_F)
\label{Eqn:10}
\end{equation}
where $WF$ is the work function of the electrode with H ion adsorption, $WF^{neutral}$ is the work function of the neutral electrode and $\Delta E_F$ is the variation of Fermi level when different amounts of electrons are introduced.The work function of the electrode is given by,
\begin{equation}
WF = E_{vac} - E_F
\label{Eqn:11}
\end{equation}
where $E_{vac}$ is the electrostatic potential of vaccum far from the electrode surface along the $c$ axis and $E_F$ is the position of the Fermi level. We have adopted this approach for calculation of $C_{redox}$.

The electric double layer capacitance (EDLC), therefore, is calculated by the relation
\begin{equation}
C_{EDL} = \frac{\epsilon_0 \epsilon_rA}{d}
\label{Eqn:12}
\end{equation}
where $\epsilon_0$ is vaccum permitivity , $\epsilon_r$ is the dielectric constant of the electrolyte. $A$ is area of the electrode and $d$ is the width of the Helmholtz layer. $\epsilon_r$ is set at 6.0 \cite{waterdiel} and $d=2.8 \r{A}$ is the radius of the Hydronium ion. 

\subsection {Structural Models}
\begin{figure}[h]
    \centering
    \includegraphics[scale=0.35]{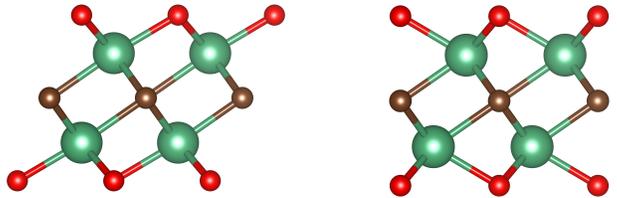}
    \caption{Structural models of $M_{n+1}C_{n}O_{2}$: (a) Oxygen atoms (red balls) occupy hollow sites A, (b) oxygen atoms occupy hollow sites B. The $M$ and $C$ atoms are denoted by green and brown balls respectively. }
    \label{fig:1}
\end{figure}

M$_{n+1}$X$_{n}$ systems are obtained from corresponding MAX compounds M$_{n+1}$AX$_{n}$ by removing the "A" layers. They consist of trilayer sheets with a hexagonal like unit cell (P6$_{3}$/mmc, space group 194). There the X layers are sand-witched between two M transition metal layers. We constructed the 2D structures by removing the 'A' element from the corresponding MAX phase systems and increasing the inter-layer distance to 15$\AA$ to avoid the interaction between the periodic images in normal directions. On the surface of the MXenes, hollow sites of two types exist: one of them is hollow site "A" for which there is no "X" atom below the "M"; the other one is hollow site "B" for which an "X" atom is available under the transition metal "M". When functional groups get attached to the MXene surfaces, they can either occupy sites "A" or sites "B" along with a possibility of a configuration where both sites are occupied. Accordingly, we constructed various structural models \cite{struct} and computed total energies of each one of them for all the systems considered. The ones with the lowest total energies were considered as the ground state configurations. We find that all the systems except $Mo_{2}C$ prefer the configuration where oxygen atoms get attached on the top of hollow sites "A" in Fig \ref{fig:1}(a). In case of $Mo_{2}C$, one oxygen is located on top of hollow site A while the other on top of hollow site B (Fig \ref{fig:1}(b)). 

We calculated the electronic structures of the MXenes considered here, by DFT based Projector Augmented Wave (PAW) method \cite{paw} as implemented in Vienna ab-initio Simulation Package (VASP)\cite{kresse1999ultrasoft}.We used a kinetic energy cut-off of 500 eV and a Monkhorst-Pack grid of 12$\times$12$\times$1 k mesh for self-consistent calculations.A larger k-mesh 42$\times$42$\times$1 was used for the calculation of the density of states(DOS).
For calculations of $C_{redox}$ and $C_{EDL}$ systems were modeled by super-cells of the basic unit cells.A 3$\times$3$\times$1 supercell and 4$\times$4$\times$1 Monkhorst-Pack  k-mesh were used for this purpose. The energy and force
convergence criteria were set at $10^{-6}$ eV and $10^{-5}$ eV/\r{A} for unit-cell calculations and $10^{-6}$ eV and 0.05 eV/\r{A} for super-cell ones.

\section{Results and Discussions}\label{results}

\begin{figure*}
 \center
 \includegraphics[scale=0.32]{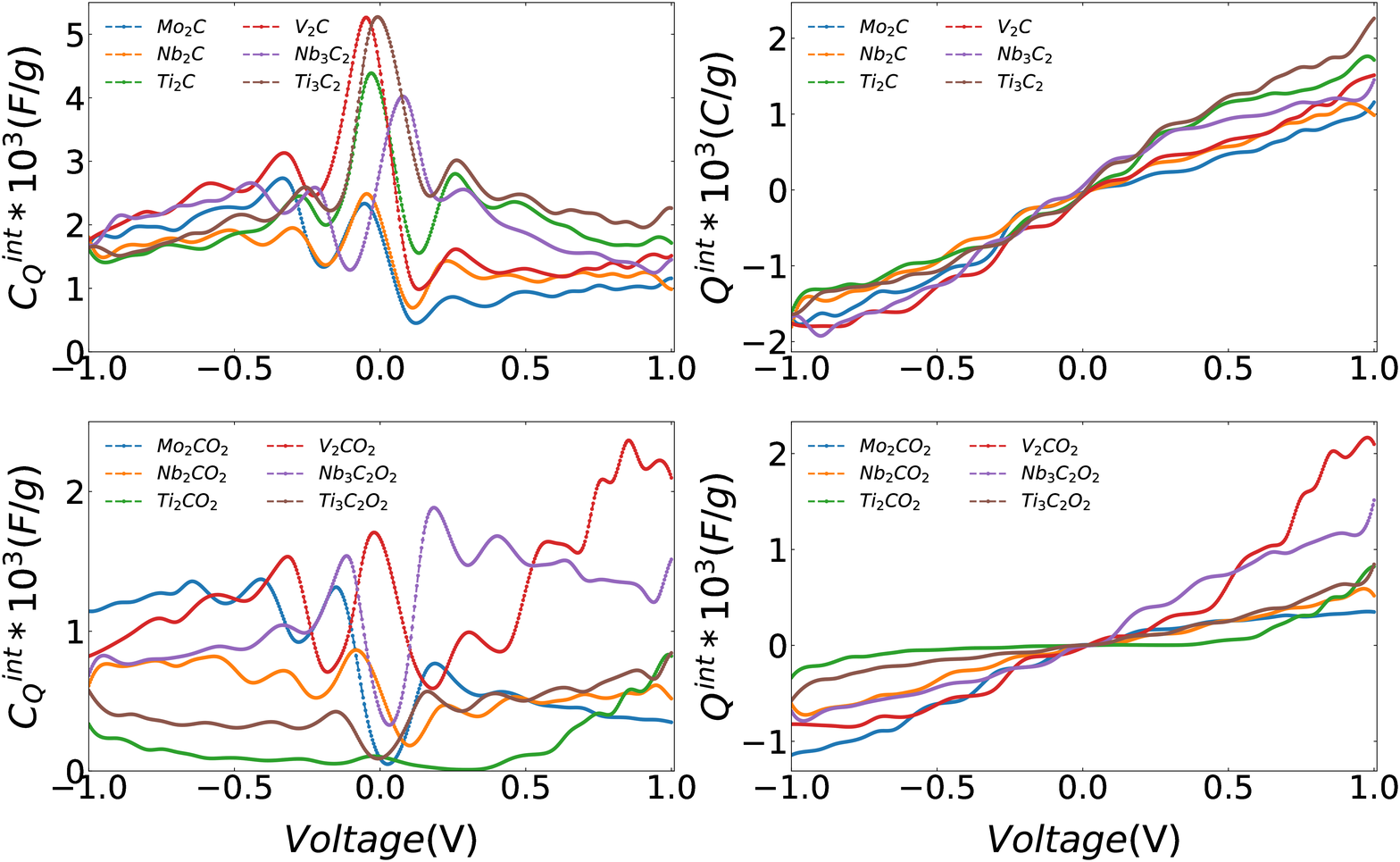}
 \caption{Variations in the Integrated Quantum Capacitance $C_Q^{int}$ (in F/g),and total Charge Storage $Q^{int}$ (in C/g) with Voltage for $M_{n+1}C_{n}$ (top Panel) and $M_{n+1}C_{n}O_{2}$ MXenes (bottom Panel)}
 \label{Fig:2}
\end{figure*}
\begin{figure}
    \centering
    \includegraphics[scale=0.12]{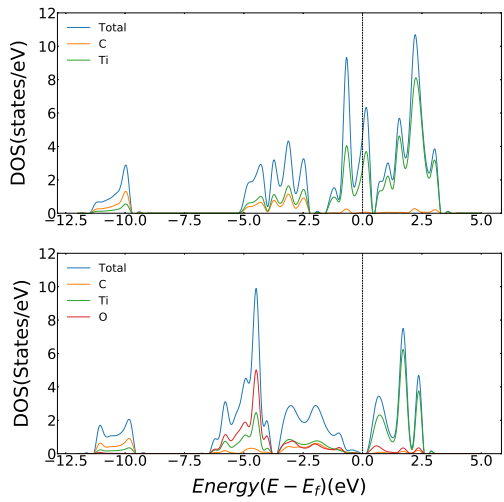}
    \caption{Partial and total densities Of States(DOS) of $Ti_2C$ (Top) and $Ti_2CO_2$(Bottom)}
    \label{Fig:3}
\end{figure}

\subsection{Quantum Capacitance}\label{R-QC}
In Figure \ref{Fig:2} we present the variations of Integrated Quantum Capacitance ($C_{Q}^{int}$) as a function of voltage. The range of voltages is selected to be $\pm 1$ Volt. The range is chosen keeping in mind that the electrochemical stability window of the electrolyte solvent is about 1.25 V at room temperature \cite{stability}. The results for pristine $M_{n+1}C_{n}$ systems by and large exhibit an uniform behaviour. Baring $Mo_{2}C$, all other systems have the maximum $C_{Q}^{int}$ near 0 V at room temperature. As the voltage is increased $C_{Q}^{int}$ decrease in a non-monotonic way with their minima around the maximum values of the voltages. Upon oxygen functionalization, the uniformity in the pattern is completely destroyed. However, the minimum $C_{Q}^{int}$ are near 0 V. As a general trend, $C_{Q}^{int}$ for functionalised MXenes are less by an order of magnitude in comparison with their pristine counterparts. These trends can be explained from the changes in the electronic structures near Fermi levels of the $M_{n+1}C_{n}O_{2}$ compounds with respect to the $M_{n+1}C_{n}$ ones. In all the cases, the $p$ states of oxygen lie deeper in the band, thus depleting the $d$ states of $M$ atoms near the Fermi levels. As an illustrative example, we show the total and partial densities of states for $Ti_{2}C$ (Figure \ref{Fig:3}). In here, the impact of functionalisation is the starkest as it opens a semiconducting gap near the Fermi level. As a result, there is significant overall reduction in $C_{Q}^{int}$ in the voltage range considered.

In Figure \ref{Fig:2} and Table \ref{Tab:1} we also depict the total charge stored ($Q^{int}$) as a function of voltage. As expected,the $Q^{int}$ are maximum at the maximum voltages. Since it depends on the capacitance, by and large, functionalised MXenes can store charge an order of magnitude less than that can be stored by the pristine ones. The trends in $Q^{int}$ with respect to materials and with respect to positive and negative voltages vary with functionalisation. In pristine compounds, while $Ti_{2}C$ and $Ti_{3}C_{2}$ can store more charge at positive voltages, the trend is different for others. In oxygen functionalised compounds, it remains the same for these two. $Nb_{2}CO_{2}$ and $Mo_{2}CO_{2}$ too follow the same trends as found in their pristine counterparts. The trends for $V_{2}CO_{2}$ and $Nb_{3}C_{2}O_{2}$, however, follow a trend different than what was found in their pristine counterparts. After functionalisation, $V_{2}C$ and $Nb_{3}C_{2}$ show significant charge storage for positive voltages while $Mo_{2}C$ has charge storage larger than the rest at negative voltages. This implies that the former two will perform better as positive electrodes while the $Mo_{2}C$ has better potential as a negative electrode, even after surface passivation.             

\begin{table}[h]
\centering
\caption{\label{Tab:1} Total Charge Storage($Q^{int}$) in C/g at $\pm$ 1.0 V for $M_{n+1}X_n$ and $M_{n+1}X_nO_2$ MXenes considered in this work.}
\resizebox{0.49\textwidth}{!}{%
\begin{tabular}{c@{\hspace{0.4cm}} cc@{\hspace{0.8cm}} cc@{\hspace{0.8cm}}}
\hline\hline
\vspace{-0.33 cm}
\\System  &    \multicolumn{4}{c}{$Q^{int}$(C/g)} \\
\hline
     & \multicolumn{2}{c}{$M_{n+1}C_n$}  & \multicolumn{2}{c}{$M_{n+1}C_nO_2$} \\
\hline
& \multicolumn{2}{c}{Voltage} & \multicolumn{2}{c}{Voltage} \\
\hline 
     &   $1$ V   &   $-1$ V               &   $1$ V    &  $\-1$ V \\
\hline

$Ti_2C$   &   1713.76  &   1621.69        &  822.65    &  335.96   \\
$V_2C$    &   1513.13  &   1758.58        &  2098.36   & 821.56    \\
$Nb_2C$   &    985.68  &   1803.79        &   518.04   & 610.39    \\
$Mo_2C$   &   1156.94  &   1677.14        &   348.61   & 1143.88   \\
\hline
$Ti_3C_2$ &   2261.77  &  1636.57         &  844.60    & 575.78    \\
$Nb_3C_2$ &   1450.72  &  1679.05         & 1514.15    & 684.81    \\
\hline
\end{tabular}
}
\end{table}

\subsection{Pseudo-Capacitance \& EDLC}\label{R-PSC}
\begin{figure}
 \center
 \includegraphics[scale=0.17]{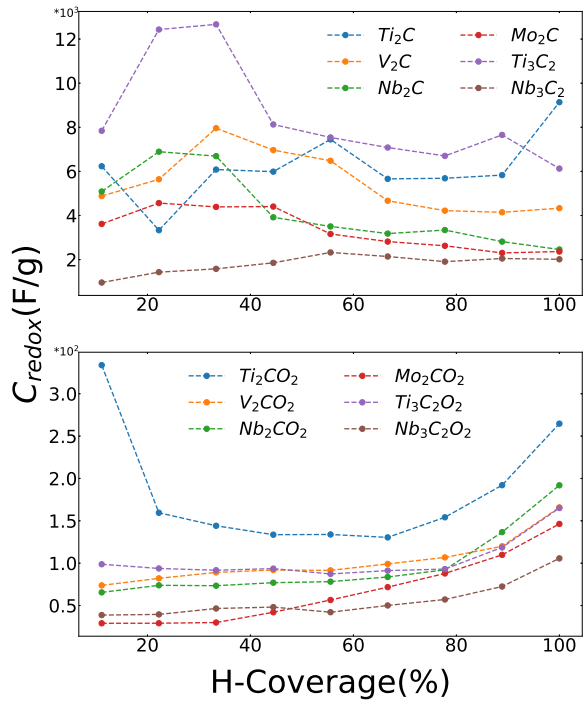}
 \caption{Variation in Pseudo-Capacitance($C_{redox}$) with H-coverage for $M_{n+1}C_n$)(top panel) $M_{n+1}C_nO_2$ (bottom panel) MXenes.}
 \label{Fig:4}
\end{figure}

 As discussed in section \ref{PSC}, Pseudo-capacitance $C_{redox}$ of the pristine and Oxygen functionalized MXene electrodes are calculated by considering them negatively charged, the electrolyte between the plates is such that  $H^+$ ions will be adsorbed on their surfaces. The continuous adsorption of $H^+$ on the electrode surfaces would lead to faradaic charge transfer and give rise to $C_{redox}$. Since the charge transfer and the work function both are supposed to vary with change in the surface coverage with $H^+$, it would be interesting to first inspect the impact on $C_{redox}$. The trends in variations of $C_{redox}$ with H-coverage for Pristine and Oxygen functionalized MXenes are shown in figure \ref{Fig:4}. Following two important trends emerge from the results: first, like $C_{Q}^{int}$,  $C_{redox}$ of pristine MXenes are one order of magnitude higher than those of oxygen functionalised MXenes, and second, $C_{redox}$ keeps on varying significantly as H-coverage changes that is as the voltage changes (the addition of each $H^+$ is equivalent to a change in the voltage). Although the variations are extremely wiggly, particularly in case of pristine MXenes, we find that in cases of oxygen functionalised MXenes, $C_{redox}^{minimum charged} < C_{redox}^{full charged}$ in general(here, $C_{redox}^{minimum charged}$ and $C_{redox}^{full charged}$ stand for minimum H-coverage and $100 \%$ H-coverage respectively) whereas $C_{redox}^{minimum charged}>C_{redox}^{full charged}$ in case of pristine compounds. Such opposing trends emerge from the trends in the charge transfer $\Delta Q$ and the changes in work function $\Delta WF$ as the H-coverage increases. While for both pristine and functionalised MXenes, $\Delta WF^{minimum charged} < \Delta WF^{full charged}$,$\Delta Q^{minimum charged} >(<) \Delta Q^{full charged}$ for pristine (functionalised) compounds. The only exception to this general trend is $Ti_2C$. For the MXenes, following the general trends, the changes in the work function have been larger than the changes in the charge transfer, making the work function the primary deciding factor in the variations of the $C_{redox}$.     

\begin{table*}
\centering
\caption{\label{Tab:2} EDLC($C_{EDL}$), Pseudo-capacitance($C_{redox}$),Quantum capacitance($C_Q$) and total capacitance ($C_T$) (in F/g units) near 0 V and -1 V are given for $M_{n+1C_n}$) and $M_{n+1}C_nO_2$) MXenes.}
\begin{tabular}{c@{\hspace{0.2cm}} cccccc@{\hspace{1.0cm}} cccccc@{\hspace{1.0cm}}}
\hline\hline
\vspace{-0.33 cm}
\\System  &    \multicolumn{6}{c}{$M_{n+1}C_n$} &  \multicolumn{6}{c}{$M_{n+1}C_nO_2$} \\
       &  $C_{EDL}$ &  $C_{redox}$ & \multicolumn{2}{c}{$C_Q$} & \multicolumn{2}{c}{$C_T$}     &  $C_{EDL}$ &  $C_{redox}$ & \multicolumn{2}{c}{$C_Q$}  & \multicolumn{2}{c}{$C_T$} \\
\hline
\ & \multicolumn{6}{c} {Voltage} &  \multicolumn{6}{c}{Voltage}\\
\hline
       &     &      &   $\sim 0$ V  &  $\sim -1$ V   &  $\sim 0$ V & $\sim -1$ V  &    &   &  $\sim 0$ V & $\sim -1$ V & $\sim 0$ V  & $\sim -1$ V \\
\hline
$Ti_2C$  &  170.55 & 9137.27  & 4243.40  &  1621.69  & 2914.63  & 1381.06  &  129.69  &  264.70  &  100.61  &  335.96  &   80.16  &  181.41 \\
$V_2C$   &  146.65  &  4331.36  &  5263.21  &  1758.58  &  2419.48  & 1262.69  & 115.77  &  165.85  & 1599.30 &  821.62  &   239.45  &  209.72  \\
$Nb_2C$  &   97.15  &  2458.69  &  2484.51  &  1803.79  &  1259.83  & 1057.47  & 83.80  &  191.00  &   811.71  &  610.39  &   265.67  & 239.78 \\
$Mo_2C$  &   83.62  &  2365.56  &  2332.59  &  1677.14  &  1194.73   &  995.46  &   69.90  &  146.33  &   432.73  &  1143.85  &  144.04  &  181.64 \\
\hline
$Ti_3C_2$  &  112.97  &   6127.86  &  4746.18  &  1636.57  &  2695.92  &  1296.56  &  91.31  &   165.11  &   147.38  &  575.78  &  93.58 &  177.41 \\
$Nb_3C_2$ &  64.32  &  2013.96  &  1794.84  &  1679.05  &  963.09  &  928.72  &  69.90  &  105.67  &  919.00  &  684.81  &   147.40  &  139.74 \\
\hline
\end{tabular}
\end{table*}

\begin{table*}
\centering
\caption{\label{Tab:3} Average Charge Transfer($\Delta Q$)($e$),Fermi-Shift($\Delta E_f$)($eV$),Change in Work-Function($\Delta WF$)($eV$) and Work-Function(WF)($eV$) for Pristine and Oxygen Functionalized MXenes.All results are for $100\%$ H adsorped system.}
\begin{tabular}{c@{\hspace{0.5cm}}cccc@{\hspace{1.0cm}}cccc@{\hspace{1.0cm}}}
\hline\hline
\vspace{-0.33cm}
\\System  &  \multicolumn{4}{c}{$M_{n+1}C_n$}  &  \multicolumn{4}{c}{$M_{n+1}C_nO_2$}\\
   & $Avg \Delta Q $ & $\Delta E_f$ &$\Delta WF$ & $WF$ & $Avg \Delta Q$ & $\Delta E_f$ & $\Delta WF$ & $WF$ \\
\hline
$Ti_2C$ & 1.622 & 0.67 & 0.693 & 4.072 & 0.499 & 1.20 & 2.565 & 1.976 \\
$V_2C$  & 1.536 & 0.70 & 1.836 & 4.961 & 0.479 & 0.515 & 3.770 & 2.168 \\
$Nb_2C$ & 1.580 & 0.84 & 0.56  & 4.225 & 0.479 & 1.21 & 2.079 & 2.387  \\
$Mo_2C$ & 1.478 & 0.95 & 0.561 & 4.604 & 0.506 & 2.11 & 2.805 & 2.421 \\
$Ti_3C_2$ & 1.614 & 0.53 & 0.518 & 4.065 & 0.463 & 1.12 & 2.680 & 2.013 \\
$Nb_3C_2$ & 1.589 & 0.74 & 0.411 & 4.294 & 0.480 & 0.715 & 2.600 & 2.121 \\
\hline
\end{tabular}
\end{table*}
 In Table \ref{Tab:2} we show $C_{redox}$ for all the MXenes considered in here, both in their pristine and oxygen-functionalised forms. The $C_{redox}$ values given are the ones for calculations with full H-coverage of the surfaces. This makes sense as the full H-coverage results are equivalent to the condition when the plates are fully charged. We find the following trends: first, the functionalised MXenes have $C_{redox}$ values one order of magnitude less than the pristine ones; second, $C_{redox}$ of 211 MXenes ($n=1$) considered here are greater than those of 321 MXenes ($n=2$).
In order to understand these trends, we first look at the trends in various quantities that make up the expression of $C_{redox}$ (Equation \ref{Eqn:10} given in Table \ref{Tab:3}.

Before analysing the results given in Tables \ref{Tab:2}-\ref{Tab:3},it is worth mentioning that the following are found:
\begin{enumerate}
    \item In case of neutral electrodes (no H-coverage), $WF^{functionalised}_{neutral} > WF^{pristine}_{neutral}$. Our calculations got this trend \cite{dipole1}right. Our calculated values of $WF^{functionalised}_{neutral} -WF^{pristine}_{neutral}$ are also in good agreement with the results in Ref \onlinecite{dipole1}. Thus, functionalisation increases the Work function of the electrodes when they are neutral.     
\item $WF$ does not vary appreciably between neutral pristine and fully charged pristine while it drops by about 3 eV in case of functionalised Ti$_{3}$C$_{2}$. The trend is similar for other MXenes. This implies that charging affects the functionalised MXenes significantly. Since a functionalised charged electrode is closer to reality than a pristine neutral one, this result is to be considered as one of the important ones and need to be understood.   
\end{enumerate}

For fully charged electrodes (Table \ref{Tab:3}), we find that across the compounds considered here, 
\begin{enumerate}
\item $\Delta E_{f}^{pristine} < \Delta E_{f}^{functionalised}$, the difference is not too big though.
\item $WF^{pristine} > WF^{functionalised}$ , the difference is about 2.0 eV.
\item $\Delta Q^{pristine} > \Delta Q^{functionalised}$ while $\Delta WF^{pristine} < \Delta WF^{functionalised}$.
\item $\Delta Q^{pristine}$ on an average is about 1.5e across compounds while $\Delta Q^{functionalised}$ on an average is 0.5e. $\Delta WF^{pristine}$ is 0.5 eV while $\Delta WF^{functionalised}$ is 2.5 eV on an average.
\end{enumerate}
The implication of the trends described above is that both charged condition and functionalisation affect the charge transfer and the changes in the Work function substantially. Therefore, explanation of the drastic reduction of $C_{redox}$ upon functionalisation would require understanding of changes in both quantities.

In case of pristine compounds, $\Delta Q$ is larger as the hydrogen gets electrons from transition metal $M$ directly while in case of functionalised one, hydrogen makes covalent bonds with oxygen and thus share electrons. The interaction is no longer directly with the transition metal but through Oxygen. Presumably Oxygen being more electronegative, gains electrons from the $M$ layers. Thus an excess of negative charge is created on the outer layer of the surface. Oxygen shares this charge with hydrogen. Hydrogen being more electropositive than Oxygen, the net charge transfer between hydrogen and functionalised substrate is much less.

It is a well known fact that the work function of a surface increases when the surface adsorbate is more electronegative. This is perfectly consistent with our calculations if we look at the work functions of the neutral electrodes. This also means that the charging brings in significant changes to reverse this trend. The question is why. 

The large $\Delta WF^{functionalised}$ indicates that there is a large change in surface dipole moment \cite{dipole1,dipole2}. This change in surface dipole depends on i) re-distribution of electronic charge between surface and adsorbate ii) dipole moment of the adsorbates and iii) change in dipole moment due to surface relaxation iv) shift in Fermi level. 
In our case more than the Fermi level shift it is the other factors that must be dominant as differences between $\Delta E_{f}^{pristine}$ and $\Delta E_{f}^{functionalised}$ are not appreciable. The charge transfer, too, changes by only 0.07 e from a neutral Ti$_{3}$C$_{2}$ to a neutral Ti$_{3}$C$_{2}$O$_{2}$,for example. However, there is a possibility of a significant charge re-distribution at the surface when an electrode is charged. To add to this, there will be a dipole contribution from OH covalent bonds. It is already been established that the internal dipole moment of OH functional group is appreciable and can have significant impact on the WF of MXenes \cite{dipole1}. Finally, there can be a contribution due to appreciable surface relaxation due to charging in case of functionalised MXenes. All these factors put together explain the trends in $C^{redox}$ found in our calculations.

Electrochemical Double Layer Capacitance($C_{EDL}$) can have a significant contribution to total capacitance. $C_{EDL}$ is calculated using Equaion \ref{Eqn:12} and presented in Table \ref{Tab:2}. In pristine compounds, $C_{EDL}$ are orders of magnitude smaller than the other contributors to the total capacitance. Upon functionalisation, the other two components reduce by orders of magnitude and thus the contributions of $C_{EDL}$ would add up significantly to the overall capacitances.

\subsection{Total capacitance $C_{T}$ and comparison with experiments}\label{R-Total}

In Table \ref{Tab:2}, the total capacitance $C_{T}$ for all $M_{n+1}C_{n}$ and $M_{n+1}C_{n}O_{2}$ compounds are presented. We have reported $C_{T}$ values at two different voltage regions: near 0 V and near the maximum that is -1 V. The reason behind choosing these two is the following: $C_{Q}$,a quantity often not taken into consideration, is dependent on the voltage directly. $C_{Q}$ is extremely important as it can reduce the $C_{T}$ substantially unless it is orders of magnitudes higher than the electrical part $C_{E}$ of the total capacitance. In our case, while calculating $C_{redox}$ we modeled the systems as negative electrodes. Therefore -1V is considered as the maximum voltage. Since the extrema in $C_{Q}$, in all cases are close to 0 and -1 V as discussed in section \ref{R-QC}, we have considered those values of $C_{Q}$ only. This way our results set the upper and lower bounds on the $C_{T}$.    

A one to one correspondence between the experimental results and our computations is not possible. The electrochemical perfomance of a supercapacitor is experimentally obtained from the capacitance versus voltage curves (cyclic voltammograms). However, the capacitance values crucially depend on the electrolyte used and the voltage scan rates. In spite of this, a rough comparison between our calculations and the available experimental results can be made. Here we compare our results with the available experimental ones by considering the experimental results that yield maximum values of the capacitances in an electrolyte providing hydrogenic ions for a negative electrode. We find that our result for $Ti_{3}C_{2}O_{2}$ agrees reasonably with the experimental value of 246 $F/g$ obtained at a scan rate of 2 mV/s in a 1 M H$_{2}$SO$_{4}$ \cite{ghidiu2014conductive}. In case of $Mo_{2}CO_{2}$, our result is in very good agreement with the experimental value of 196 $F/g$ obtained at a scan rate of 2 mV/s in a 1 M H$_{2}$SO$_{4}$ \cite{halim2016synthesis}. The experimental value of 51 F/g in $Ti_{2}CO_{2}$ obtained at a scan rate of 5 mV/s in a KOH solution \cite{ti2c2015} too is in fair agreement with the minimum $C_{T}$ reported by us. In case of $V_{2}C$, the degree of agreement between the experiments and our calculations varied depending upon the electrolyte used in the experiment. The maximum capacitances obtained in experiments were 164 F/g \cite{v2c2020},184 F/g, 225 F/g, and 487 F/g \cite{shan2018two} in Na$_{2}$SO$_{4}$,KOH, MgSO$_{4}$ and H$_{2}$SO$_{4}$ solutions. Our calculated result is closer to the first three experimental values.

 \section{Summary}
 Mxenes are the new class of low-dimensional materials with tremendous potentials as supercapacitors. Although several MXenes have been synthesised and their electrochemical characteristics studied in recent past, to our knowledge, there is no systematic investigations into their capacitances by analysis of the various contributions coming from different sources. This work is the first in that direction where the understanding of their capacitative behaviours is done from the electronic structures. The fairly good agreement between our results and the experimental ones for the systems where experiments have been done demonstrate the correctness of our approach and that the models used closely mimic the experimental situations. Important aspects like the quantum capacitances and their variations with voltages, the impact of charging the electrodes and the surface passivation have been discussed in detail. The results also establish the following: first, the results of pristine and the oxygen-functionalised ones provide the limiting values of the total capacitance. Pristine compounds which exhibit order of magnitude larger values of the capacitances are idealised systems while fully surface passivated ones are the other extremes. In an experiment, the surface may not be fully functionalised or there may be mixture of functional groups. Thus the truth, in all likelihood, will be in between the two extremes. Therefore, all the compounds considered here will display excellent capacitative behaviour in reality if they are used as negative electrodes; second, although the transition metal constituents in the compounds considered are chosen from both $3d$ and $4d$ series, there is hardly any noticeable difference or a trend in their relative behaviours as far as the capacitances are considered; third, we find $Nb_{n+1}C_{n}$ as another potential negative electrode that can be used in a supercapacitor. It's $C_{T}$ value being close to the other compounds, should attract the experimentalists to assess it's electrochemical performance.




\section{Acknowledgement}
The authors gratefully acknowledge the Department of Science and Technology, India, for the computational facilities under Grant No. SR/FST/P-II/020/2009 and IIT Guwahati for the PARAM supercomputing facility. Useful discussions with Dr. U.N.Maiti are also acknowledged.

\end{document}